\input{aipcheck}

\documentclass[
    ,final            
  ]
  {aipproc}

\layoutstyle{6x9}

\begin{document}

\title{Multi-Lepton Events at H1 and Search for Doubly-Charged Higgs Bosons}


\author{Andr\'e Sch\"oning}{
  address={ETH H\"onggerberg Z\"urich, Switzerland}
}

%

\begin{abstract}

 Events with two or more leptons (electrons or muons) with high transverse 
momentum are measured in electron-proton collisions at HERA using the 
data sample collected in the period 1994-2005. 
Multi-lepton events at high transverse momenta are of special interest as
these signature might reveal new physics beyond the Standard Model.
An example is 
the single production of doubly-charged Higgs bosons $H^{\pm \pm}_{L,R}$, which
couple to leptons of the $i^\prime$th and $j^\prime$th generation via 
Yukawa couplings $h^{L,R}_{ij}$.
Results from a search for doubly-charged Higgs bosons in the decays
into electrons, muons and taus are presented using data taken in the period 
1994-2000. 
No evidence for doubly-charged Higgs production is found and we derive limits 
on the $h_{ee}^{L,R}$ and $h_{e \mu}^{L,R}$ Yukawa couplings as a function of 
the $H^{\pm\pm}_{L,R}$ mass.

\end{abstract}

\maketitle


\section{Introduction}

The H1 experiment has reported an excess of di-electron and 
tri-electron events with invariant masses above 100~GeV observed in $ep$
collisions at HERA~I in the period 1994-2000 \cite{me}.
In this talk results from a recent multi-lepton search at high 
transverse momentum 
are presented  using the HERA~I+II data samples collected in 1994-2005. 

Multi-lepton events at high transverse momenta are of special interest as
this signature might reveal new physics beyond the Standard Model.
An example is 
the single production of doubly-charged Higgs bosons $H^{\pm \pm}_{L,R}$, which
couples to leptons of the $i^\prime$th and $j^\prime$th generation via 
Yukawa couplings $h^{L,R}_{ij}$.

Doubly-charged Higgs bosons ($H^{\pm \pm}$) appear naturally
in various extensions of the Standard Model (SM)
in which the usual Higgs sector is extended by one or more
triplet(s)~\cite{HTM,pati}.
Examples are provided by some Left-Right Symmetric (LRS)
models~\cite{moha3,moha4} with 
a spontaneously broken $SU(2)_L \times SU(2)_R \times U(1)_{B-L}$ symmetry.
These models are of particular interest as Higgs triplets can give 
Majorana masses to neutrinos, which are known to be massive from recent
experimental data.
Results from a search for doubly-charged Higgs bosons in the decays
into electrons, muons and taus are presented using data taken in the period 
1994-2000. 

\section{Events with Multi-Leptons} 
\begin{figure}
   \begin{picture}(400,360)
   \put(0,185) {
  \put(0,0){\includegraphics[height=.25\textheight]{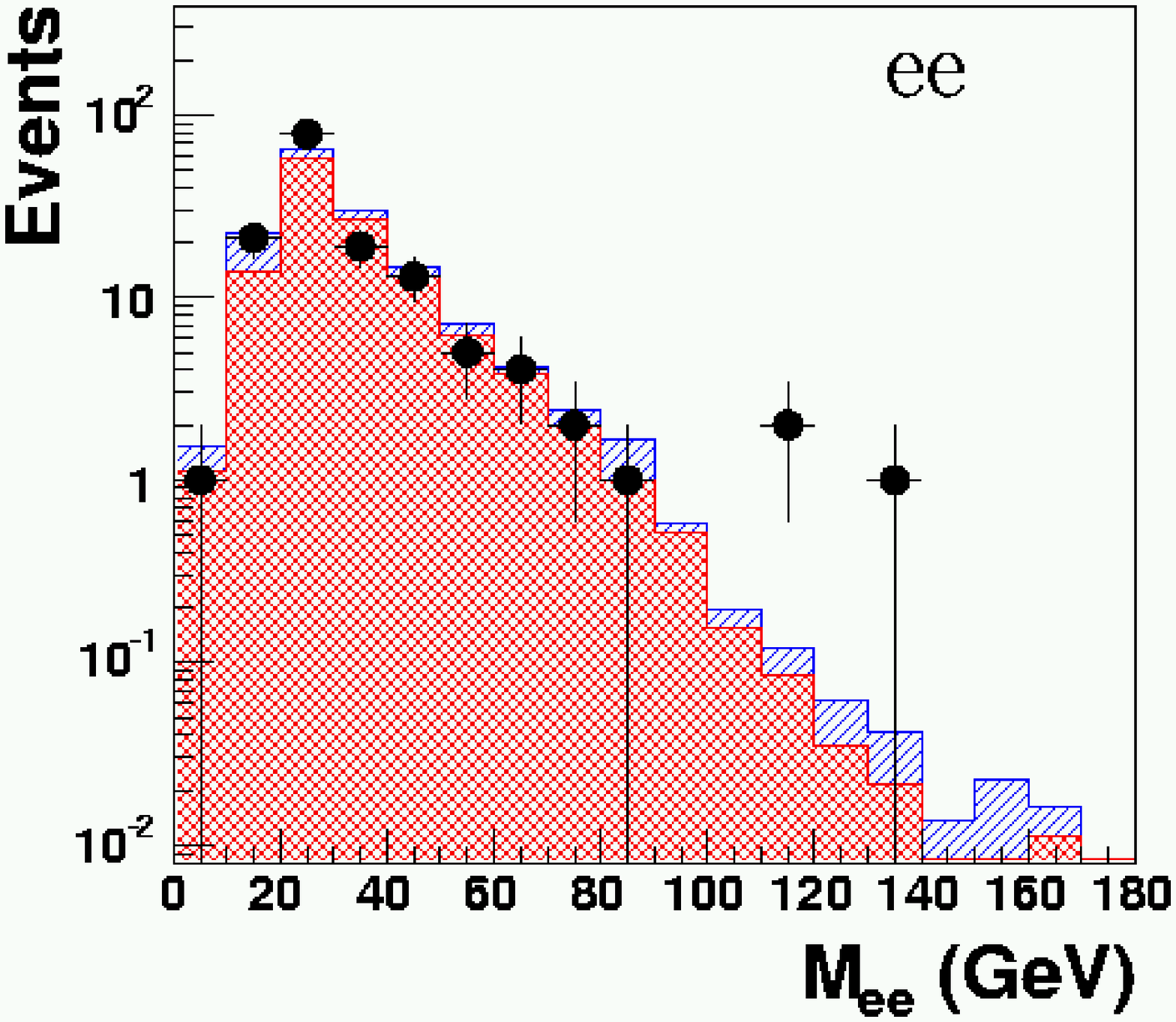}}
  \put(200,0){\includegraphics[height=.25\textheight]{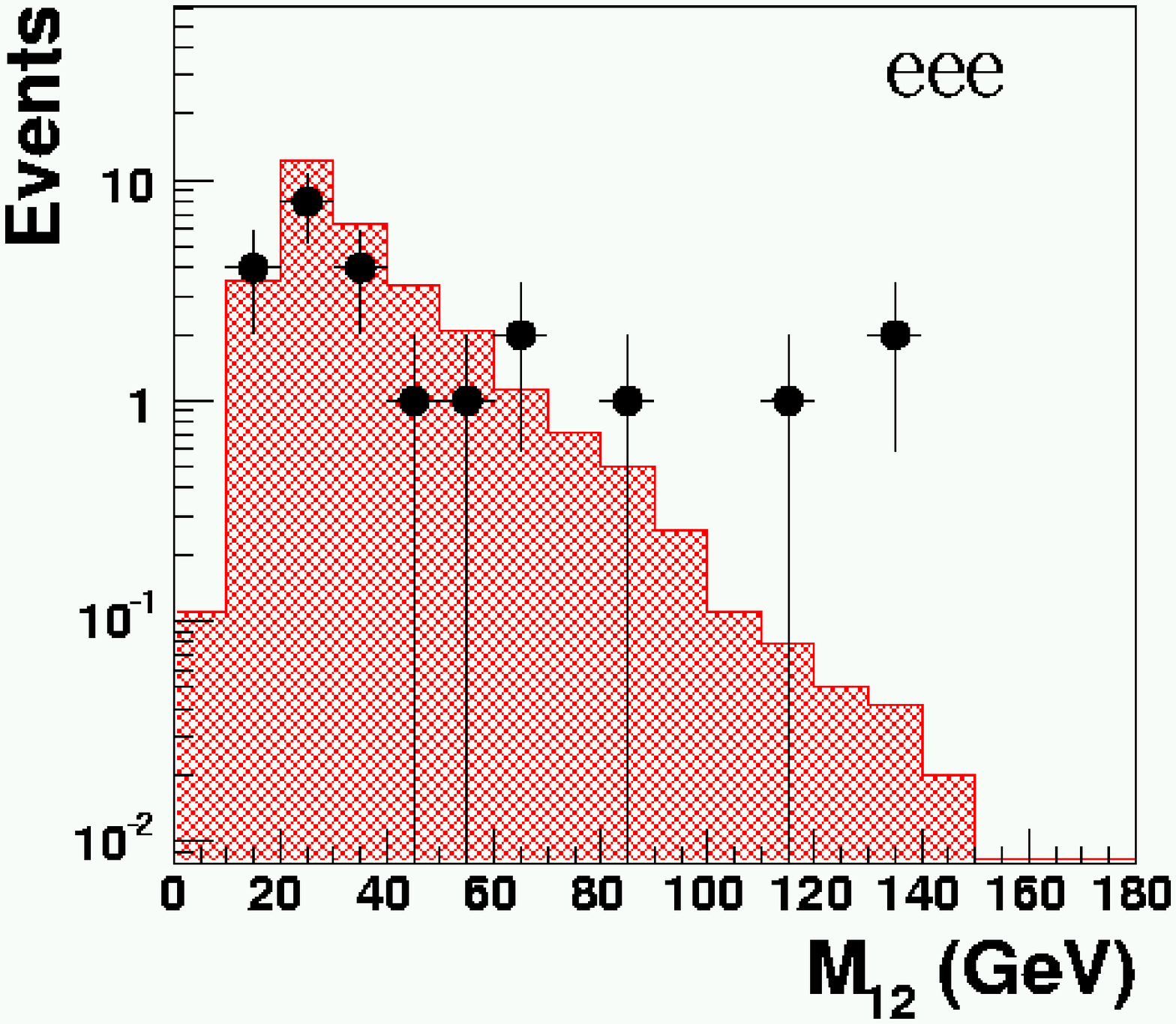}}
  \put(92,110){\includegraphics[height=.038\textheight]{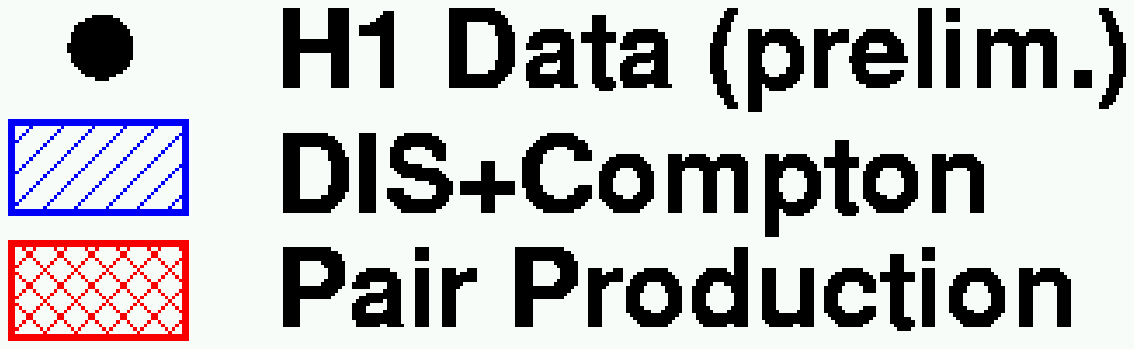}}
  \put(100,163){H1 Preliminary HERA I+II $e^\pm p$ (163~pb$^{-1})$}
}

%
  \put(0,0){\includegraphics[height=.25\textheight]{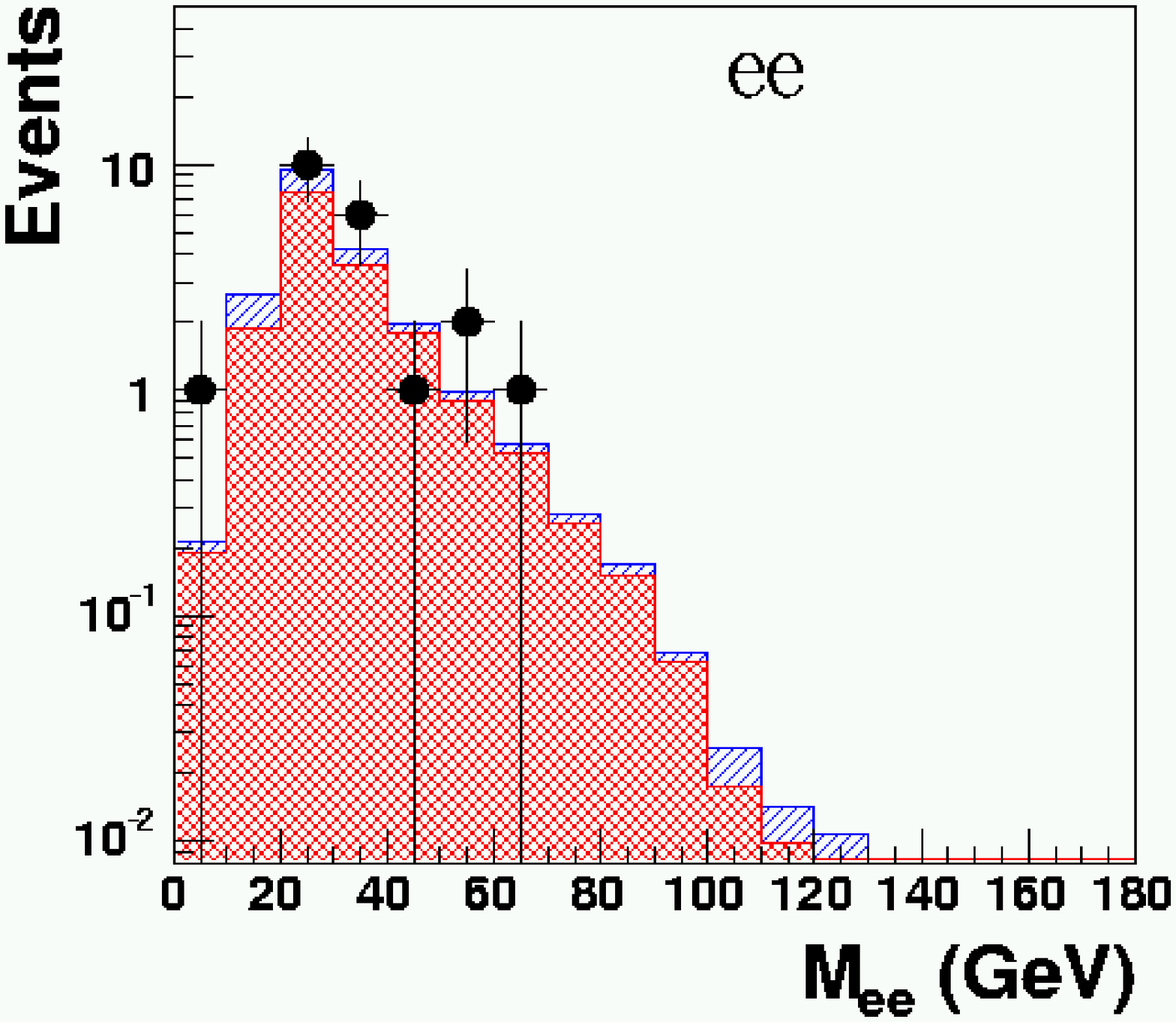}}
  \put(200,0){\includegraphics[height=.25\textheight]{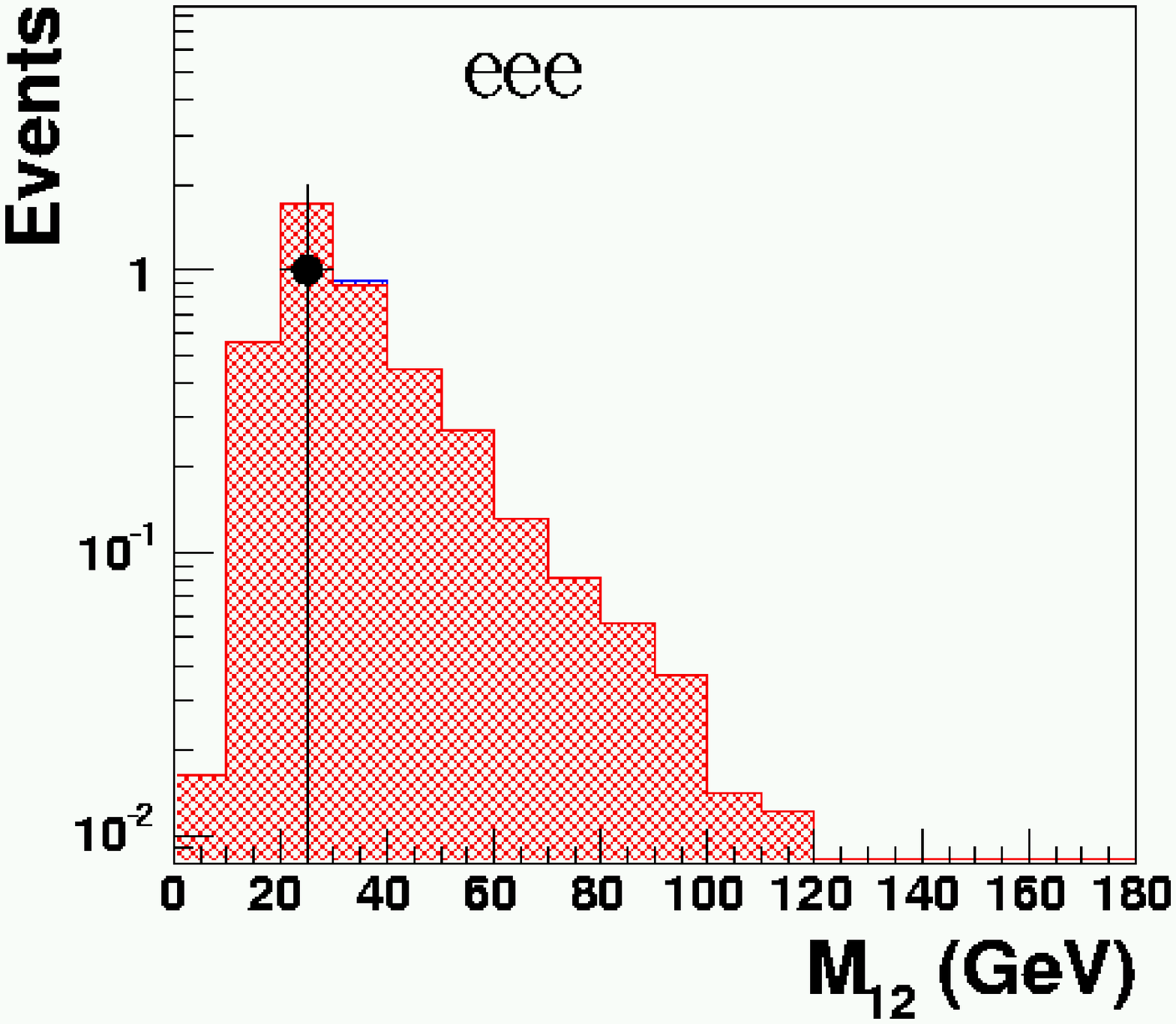}}
  \put(85,110){\includegraphics[height=.038\textheight]{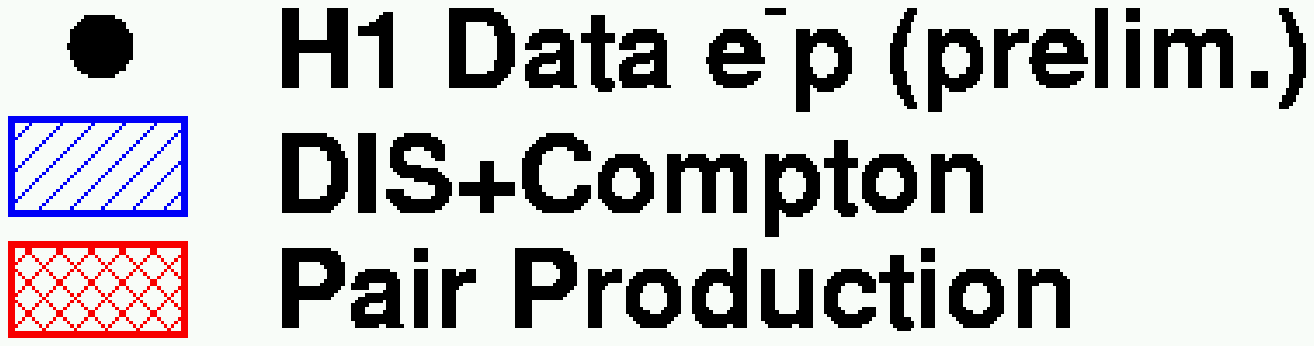}}
  \put(100,163){H1 Preliminary HERA II    $e^- p$ (21~pb$^{-1})$}

   \end{picture}
  \caption{Invariant masses $M$ of the two highest $p_T$  leptons
  compared to expectations for $ee$ and $eee$ classified event.
  The data samples  correspond to an integrated luminosity of 
  163~pb$^{-1}$ (1994+2004, top figures)
  and to 21~pb$^{-1}$ of
  the recent $e^-p$ data taken in 2004/05 (bottom).} 
  \label{fig:mass}
\end{figure}

Events with clearly identified isolated electrons and muons at high 
transverse momentum ($p_T$) 
are selected and classified into the following final state topologies:
$ee$, $eee$, $e\mu$, $\mu\mu$, $e\mu\mu$, etc.). 
Within the Standard Model such events are mainly produced
via photon-photon collisions. 
The invariant mass distribution of the two
highest $p_T$ electrons are shown for the $ee$ and $eee$ classes in 
Fig.~\ref{fig:mass} using 163~pb$^{-1}$ of HERA~I+II data. 
The event yields for all classes after the cut 
$M_{\ell{\ell^\prime}}>100$~GeV are shown in Tab.~\ref{tab:epos}.
An excess of data events over the SM expectation 
is seen in the $ee$ and $eee$ classes.

For some event classes the same analysis was repeated using the most recent
$e^-p$ dataset from 2004/05 (21~pb$^{-1}$), see Fig.~\ref{fig:mass}. 
No new event was found for  $M_{\ell{\ell^\prime}}>100$~GeV in the analysed
$ee, e\mu, eee, e\mu\mu$ event classes in agreement with the SM expectation.

\begin{table}
\begin{tabular}{lccc}
\hline
channel \quad ~ & ~ \quad data events \quad ~ & ~ \quad Standard Model \quad ~
& ~ \quad Pair Production \quad ~ \\ \hline
$ee$          & 3 & $0.4  \pm 0.1$    & $0.32$ \\
$\mu \mu$     & 0 & $0.04 \pm 0.02 $  & $0.04$ \\
$e \mu$        & 0 & $0.31 \pm 0.03 $  & $0.31$ \\ \hline

$eee$         & 3 & $0.31 \pm 0.08 $  & $0.31$ \\
$e\mu\mu$ \ $(M_{e\mu})$& 1 & $0.04 \pm 0.01 $  & $0.04$ \\
$e \mu \mu$ \ $(M_{\mu\mu})$ & 1 & $0.02 \pm 0.01 $  & $0.02$ \\
\hline
\end{tabular}
\caption{H1 preliminary event yields for 1994-2004 $e^\pm p$ data (L=163~pb$^{-1}$) of the different 
multi-lepton channels for
$M_{ll^\prime}>100~GeV $. The numbers are compared with the expectations from the Standard Model and from pair production only.}
\label{tab:epos}
\end{table}

\section{Doubly-Charged Higgs Production}
The search for single doubly-charged Higgs production is based on studies
of dilepton production $ee$, $\mu\mu$ \cite{me,mm}, 
and new multi-lepton searches for the $e\mu$ and
$\tau\tau$ final states at high $p_T$. 
The analyses use data collected 
at HERA~I from 1994-2000 ($ee$, $\mu\mu$ and $e\mu$) and 1999/2000
($\tau\tau$).
The selections involving electrons and muons are identical to those described in
the previous section.
Tau pairs are searched for taking into account leptonic and hadronic $\tau$-decays
in the event classes $\tau\tau \rightarrow e\mu, ej, \mu j, jj$ 
plus missing energy from the unobserved tau neutrinos.
The invariant mass of the $H^{\pm \pm}$ is fitted by exploiting energy and
momentum conservation resulting in a resolution of $2.5-4.0$~GeV.

\begin{figure}
  \includegraphics[height=.27\textheight]{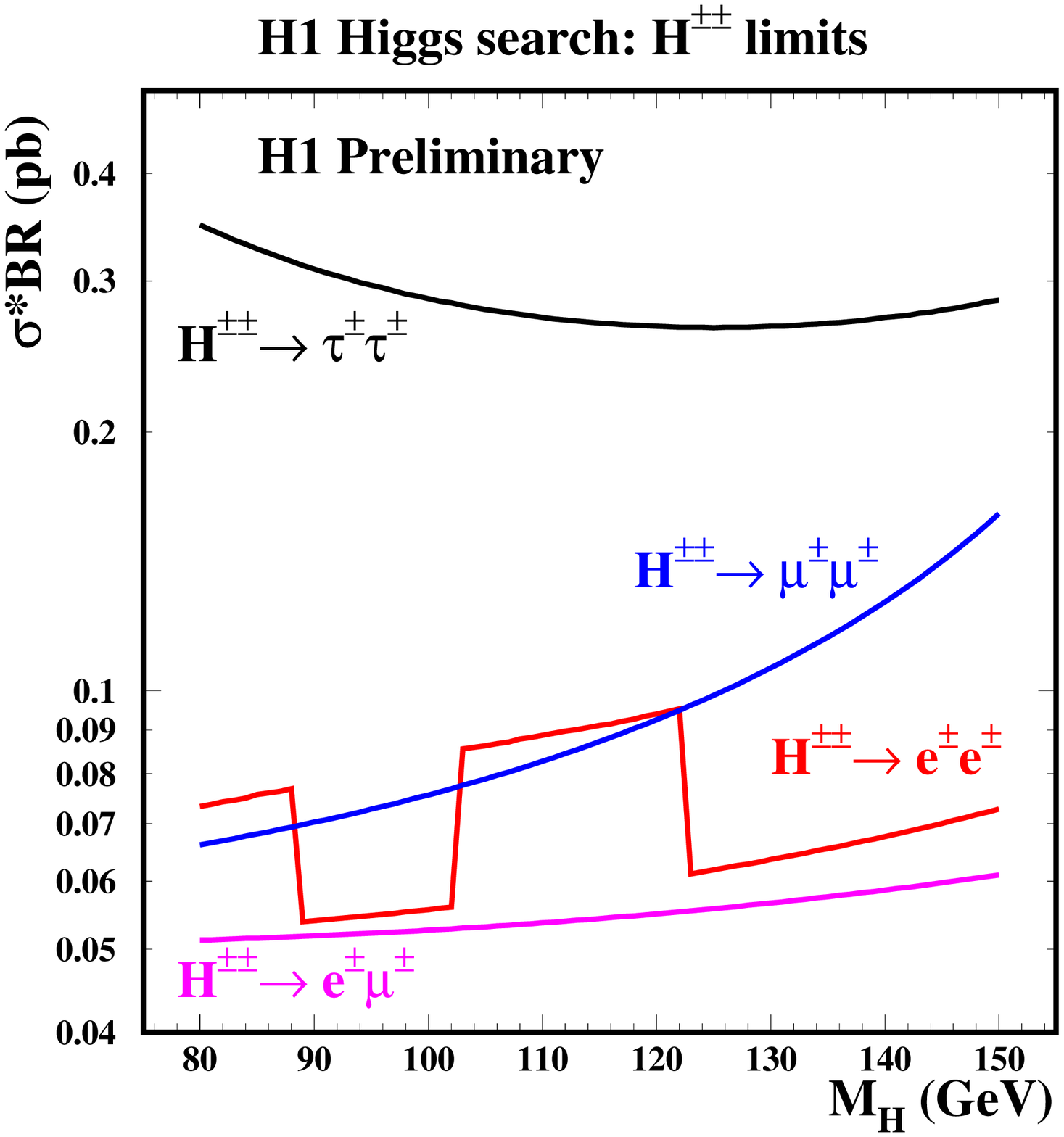}
  \includegraphics[height=.27\textheight]{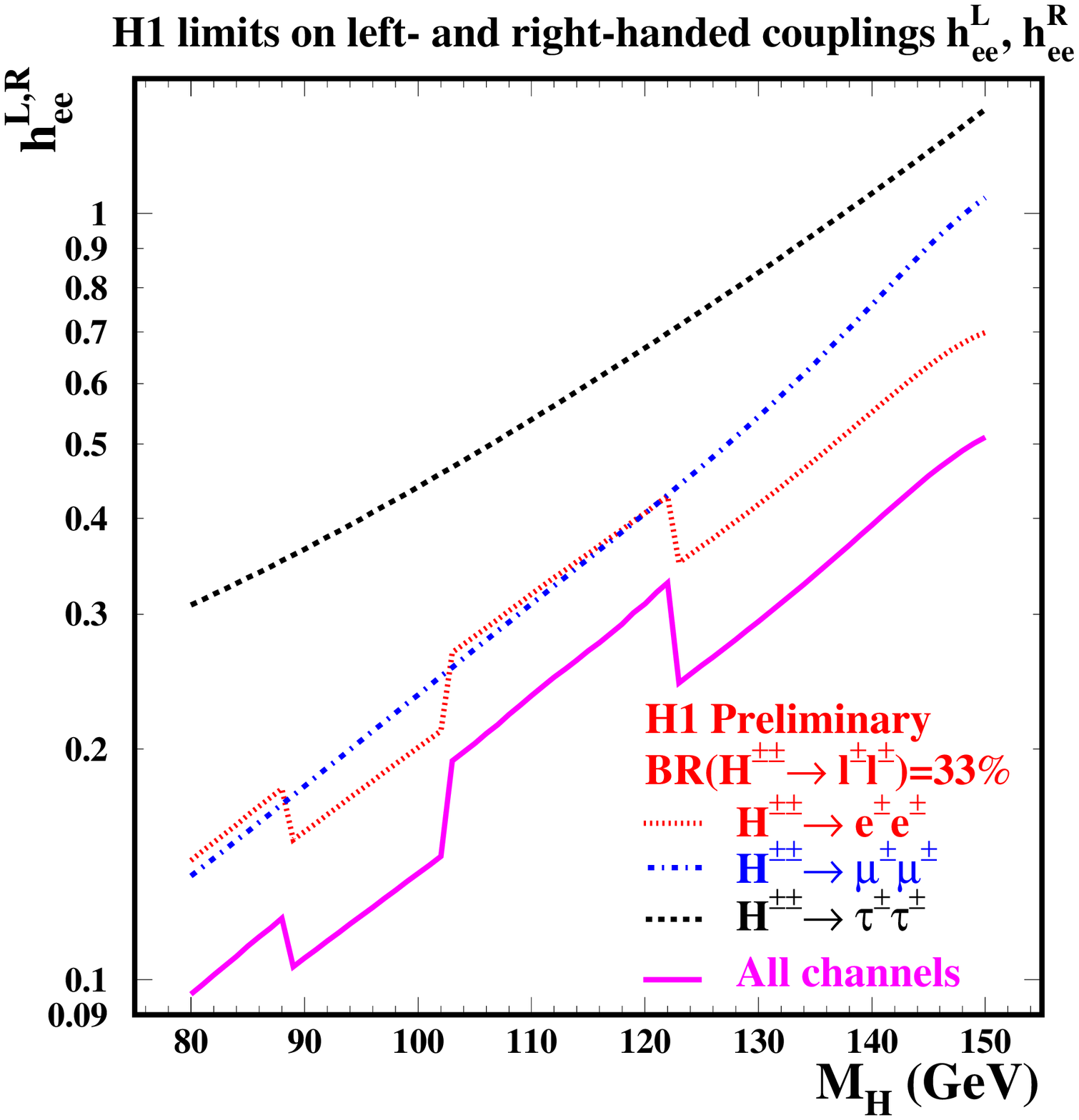}
  \caption{Left: upper limits at 95\% CL on 
  $\sigma 
  (e^\pm p \rightarrow e^\mp H^{\pm\pm} X) \ \times \
   {\rm BR}(H^{\pm\pm} \rightarrow \ell^\pm \ell^{\prime
   \pm})$ as a function of the doubly-charged Higgs mass.
  Right: exclusion limits on the coupling $h_{ee}^{L,R}$ at 95\% CL as function
  of $M_H$ for the decay channels $H^{\pm\pm} \rightarrow e^\pm e^\pm, 
  \mu^\pm \mu^\pm, \tau^\pm \tau^\pm$ and their combination (full curve)
  assuming democratic couplings, ie. ${\rm BR}(H^{\pm\pm} \rightarrow \ell^\pm
  \ell^\pm)=1/3$.}
\label{fig:sbr}
\end{figure}

\begin{figure}
  \includegraphics[height=.27\textheight]{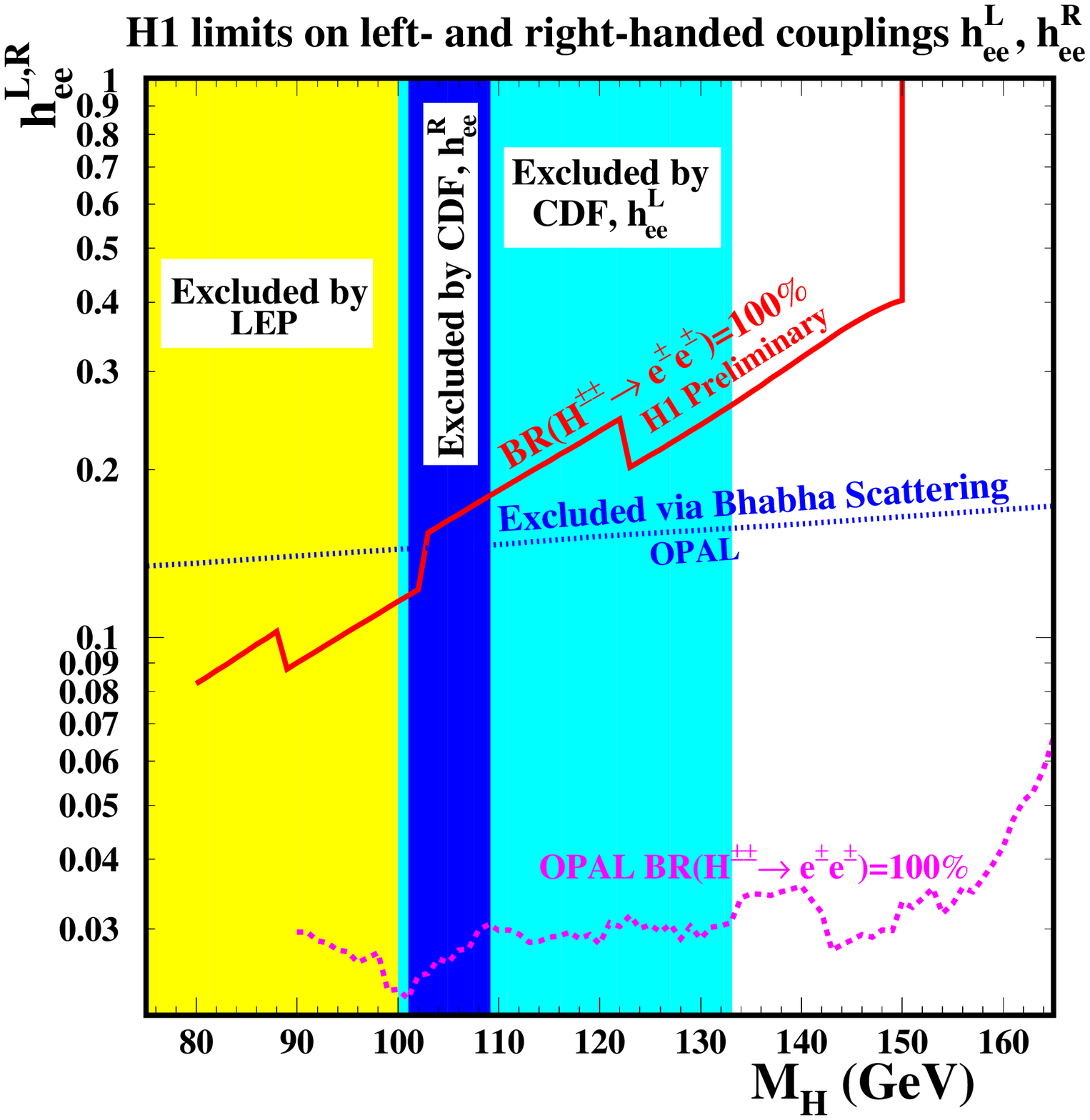}
  \includegraphics[height=.27\textheight]{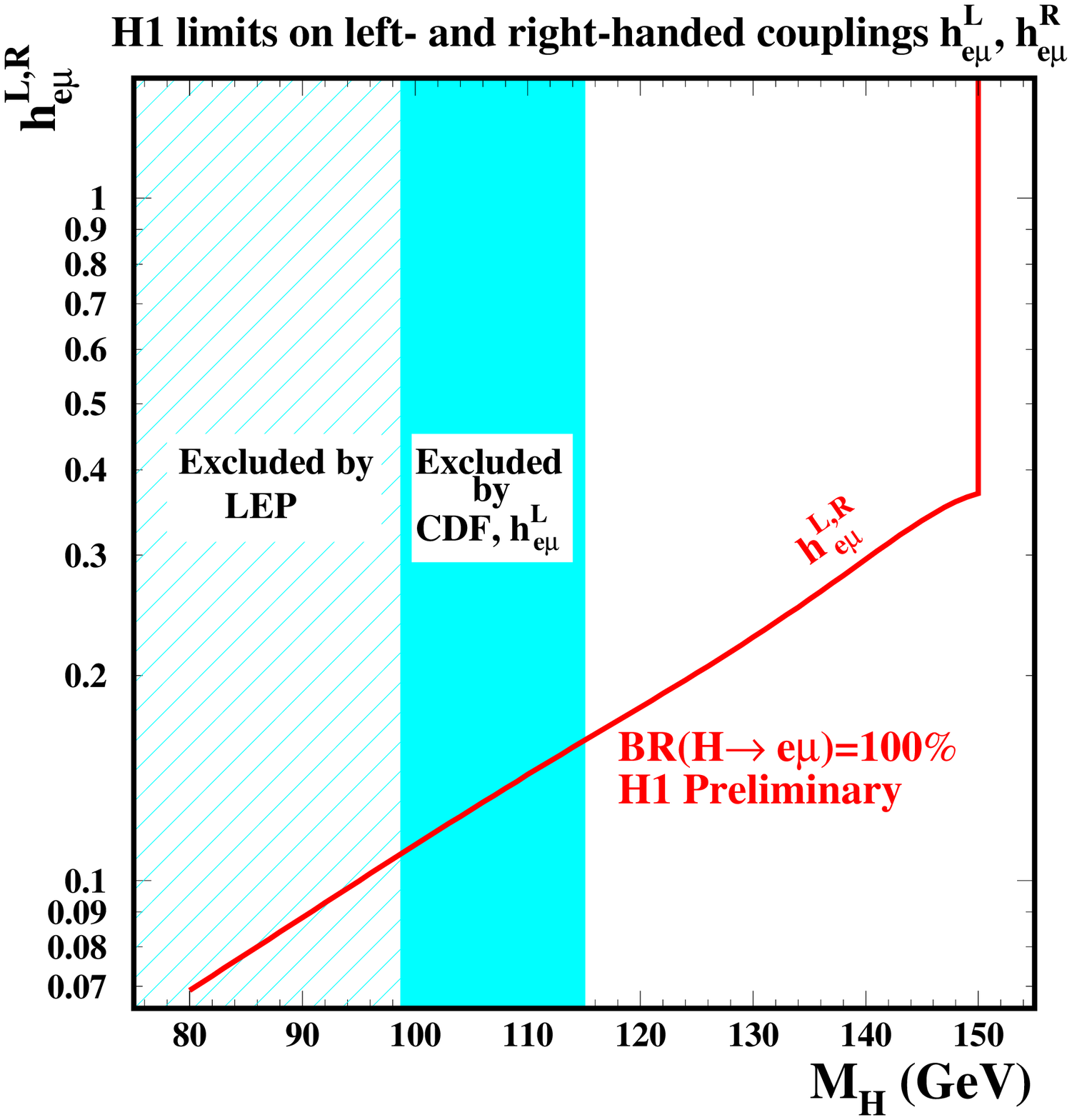}
  \caption{Left: exclusion limits on the coupling $h_{ee}^{L,R}$ at 95\% CL as
  a function of the doubly-charged Higgs mass for the decay 
  $H^{\pm\pm} \rightarrow e^\pm e^\pm$ assuming 
  ${\rm BR}(H^{\pm\pm} \rightarrow e^\pm e^\pm)=1$. 
  Right: exclusion limits on the coupling $h_{e\mu}^{L,R}$ at 95\% CL as
  a function of $M_H$ for the decay 
  $H^{\pm\pm} \rightarrow e^\pm \mu^\pm$ assuming 
  ${\rm BR}(H^{\pm\pm} \rightarrow e^\pm \mu^\pm)=1$. 
The results are compared to direct and indirect limits (Bhabba scattering)
obtained by OPAL (single production), and LEP and CDF (pair production).}
\label{fig:coupl}
\end{figure}

Further $H^{\pm \pm}$ selection criteria are applied to all multi-lepton final
states. Events
where the charges of the highest $p_T$ leptons are measured to be inconsistent 
with the $H^{\pm \pm}$ hypothesis are rejected. 
For the $H^{\pm \pm} \rightarrow e^\pm e^\pm$
decay both electrons are required to have large transverse energies in the
calorimeter. After the final selections only one candidate event 
($H^{\pm \pm} \rightarrow e^\pm e^\pm$)
is found for $M_H>100$~GeV.
Various limits are derived as function of $M_H$
at 95\%~CL on the product $\sigma \times {\rm BR}$
 for each decay topology, see Fig~\ref{fig:sbr}~left, and on the
coupling $h_{ee}^{L,R}$ assuming a model with democratic couplings 
${\rm BR}(H^{\pm\pm} \rightarrow  l^{\pm}l^{\pm}) = 1/3$ (right).
In Fig.~\ref{fig:coupl} exclusion limits on $h_{ee}^{L,R}$ (left) and 
$h_{ee}^{L,R}$ (right)
are shown as function of $M_H$ assuming a decay branching ratio of 100\% for
each case.
For comparison direct and indirect limits from other experiments are shown 
\cite{opal-singleprod,leppairprod,cdf-pairprod}.

\section{Conclusion}
Multi-lepton events with electrons and muons are studied using 
HERA I+II data.
No new di-electron and tri-electron events are observed 
with invariant masses above 100~GeV
in the most recent $e^-p$ dataset.
 More integrated luminosity is required to enhance the sensitivity
at high masses.
No evidence for doubly-charged Higgs production is found and various limits 
on $H^{\pm\pm}_{L,R}$ Yukawa couplings $h_{\ell \ell^\prime}^{L,R}$ 
were derived.







\bibliographystyle{aipproc}   

\bibliography{sample}



\end{document}